\documentclass[preprint]{aastex}
\usepackage{epsfig}

\usepackage{natbib}
\usepackage{subfigure}
\usepackage{multirow}
\usepackage{amssymb}
\usepackage{captcont}

\newcommand{\msun}{\mbox{M$_{\odot}$}}

\begin{document}

\title{Evolution of the Stellar Mass Tully-Fisher Relation in Disk Galaxy Merger Simulations}

\author{Matthew D. Covington\altaffilmark{1} }
\affil{Physics Department, University of California,
    Santa Cruz, CA 95064}
\email{covin039@umn.edu}

\author{Susan A. Kassin}
\affil{Sub-Department of Astrophysics, University of Oxford, Oxford OX1 3RH, UK
}
\email{s.kassin1@physics.ox.ac.uk}

\author{Aaron A. Dutton} 
\affil{UCO/Lick Observatory and Department of Astronomy, University of
  California, Santa Cruz, CA 95064}

\author{Benjamin J. Weiner}
\affil{Steward Observatory, University of Arizona, Tucson, AZ 85721}

\author{Thomas J. Cox}
\affil{Institute for Theory and Computation, Harvard-Smithsonian Center for Astrophysics, Cambridge, MA 02138}

\author{Patrik Jonsson}
\affil{Santa Cruz Institute of Particle Physics, Santa Cruz, CA 95064}

\author{Joel. R. Primack\altaffilmark{1}}
\affil{Physics Department, University of California,Santa Cruz, CA 95064}

\author{Sandra M. Faber}
\affil{UCO/Lick Observatory and Department of Astronomy, University of
  California, Santa Cruz, CA 95064}

\author{David C. Koo}
\affil{UCO/Lick Observatory and Department of Astronomy, University of
  California, Santa Cruz, CA 95064}


\altaffiltext{1}{Santa Cruz Institute for Particle Physics, Santa Cruz, CA 95064}

\begin{abstract}
There is a large observational scatter toward low velocities in the
stellar mass Tully-Fisher relation if disturbed and compact objects
are included. However, this scatter can be eliminated if one replaces
rotation velocity with $\rm S_{\rm 0.5}$, a quantity that includes a
velocity dispersion term added in quadrature with the rotation
velocity. In this work we use a large suite of hydrodynamic N-body
galaxy merger simulations to explore a possible mechanism for creating
the observed relations.  Using mock observations of the simulations,
we test for the presence of observational effects and explore the
relationship between $\rm S_{\rm 0.5}$ and intrinsic properties of the
galaxies.  We find that galaxy mergers can explain the scatter in the
TF as well as the tight $\rm S_{\rm 0.5}$-stellar mass relation.
Furthermore, $\rm S_{\rm 0.5}$ is correlated with the total central
mass of a galaxy, including contributions due to dark matter.

\end{abstract}


\section{Introduction}

One of the goals of a theory of galaxy formation is to reproduce the
scaling relations of both early and late type galaxies.  There has
been significant work on understanding galaxy formation from analytic,
semi-analytic, and numerical perspectives dating back to the
collapsing gas cloud model of \citet{egge}.  For early type galaxies,
scaling relations between their sizes, luminosities or stellar masses,
and velocity dispersions form a two-dimensional plane called the
``Fundamental Plane" \citep{dress,djor}.  For late type galaxies, a
simple plane does not exist \citep{kappa}, and the relation between
luminosities or stellar masses and rotation velocities \citep[i.e.,
  the Tully-Fisher (TF) relation;][]{TF} is independent of size and
surface brightness \citep[e.g.,][]{cour}.

The Fundamental Plane was found to be a consequence of the virial
theorem, if one assumes a constant mass-to-light ratio and homologous
mass structure \citep{fabe,djor,bend,pahr}.  However, the situation is
more complicated since there is a tilt observed between the
Fundamental Plane and the plane resulting from the virial theorem.
This tilt is likely due to a combination of stellar population effects
and another factor which has yet to be determined
\citep[e.g.,][]{pahr,gerh,bert,hopk}.  \citet{hopk} argue that this
other factor results from a varying ratio of total enclosed mass
within the effective radius ($R_e$) to stellar mass.  This ratio must
depend on mass such that lower mass ellipticals are more
baryon-dominated within $R_e$ than their higher mass counterparts.

For disk galaxies, the scaling relations between velocity, luminosity
or stellar mass, and size can broadly be understood by semi-analytic
galaxy formation models which are based upon the dissipationless
collapse of a cold dark matter halo and angular momentum conservation
\citep[e.g.,][]{momao,vdb01,dalc,dejo,dut7}.  Such models have their
roots in the analytic formulations of \citet{whit} and \citet{fall}
which in turn have their roots in the works of \citet{silk} and
\citet{rees}.  A detailed prediction of disk galaxy scaling relations,
which necessitates simultaneous matching of the zero-points of the TF
relation, galaxy sizes, and the luminosity/stellar mass functions has
proven elusive \citep[e.g.,][]{bens,dut7}.  Models can be constructed
that reproduce these properties \citep{dut9}, but they require
efficient mass outflows (a.k.a. feedback), and the absence of dark
halo adiabatic contraction \citep{blum}.  On the purely numerical
front, hydrodynamic + N-body codes have been able to form disk-like
systems only under controlled conditions and for halos carefully
selected based upon their merger histories \citep[e.g.,][]{gov4, gove,
  rob4, scan}.  Furthermore, the formation of realistic bulge-less
galaxies remains an unsolved problem for cosmological simulations
\citep[e.g.,][]{maye}.

Cosmological hydrodynamical simulations have been able to form
galaxies with disks that reproduce the slope of the TF relation
(e.g. Navarro \& Steinmetz 2000; Eke et al. 2001, Portinari et
al. 2007, Governato et al. 2007; Piontek \& Steinmetz 2009).  However,
as with analytic and semi-analytic models, reproducing the TF zero
point, and observed disk sizes has been a challenge.  If the rotation
velocity is measured at a radius which encloses most of the baryons,
then simulated galaxies can reproduce the zero point of the TF
relation (e.g. Governato et al. 2007). However, if the rotation
velocity is measured at a smaller radius (such as 2.2 K-band disk
scale lengths), the same simulations of Governato et al. (2007)
over-predict the rotation velocities (Dutton \& Courteau 2008).  This
failure is due to the simulations having the wrong distribution of
baryons and/or dark matter within the optical parts of galaxies.

In summary, although the scaling relations of early and late type
galaxies can be explained in broad terms, detailed predictions have
yet to fully mature.  In addition, since early type galaxies are
generally expected to form through mergers of late type systems, the
scaling relations of early types should therefore descend from those
of late types.  There has been work with simulations of mergers of
late types to try to understand this \citep[e.g.,][]{RobertsonFP,
  remnants,hop9}.  \citet{Dekel06}, \citet{hopk}, and Covington et
al. (in prep) in particular argue that late type scaling relations may
evolve into those of early types if the varying gas content of disk
galaxies with mass and dissipation during mergers are taken into
account in models.  However, not all galaxies in today's universe are
the well-ordered disk or elliptical systems for which most theories
make predictions.  Many galaxies have disturbed or irregular
morphologies and/or kinematics, and the fraction of these galaxies
increases with redshift to at least $z\sim1$ \citep[see e.g.,][for
  morphologies and kinematics, respectively]{vand,Kassin}.  With this
in mind, we examine the findings of \citet{Kassin} which attempt to
include such irregular systems in a study of the Tully-Fisher scaling
relation, and to link early and late types through their kinematics.

In particular, we focus on the scaling laws that relate galactic
orbital speeds to stellar masses: the TF relation \citep{TF} for
rotating disk galaxies, which relates stellar mass to rotation
velocity (${\rm V_{rot}}$), and the Faber-Jackson relation (FJ)
\citep{FJ} for spheroidal (elliptical) galaxies, which relates stellar
mass to velocity dispersion ($\sigma$). The conventionally assumed
pathway for forming ellipticals is through mergers of disk galaxies
\citep{Barnes92}, in which case the FJ relation descended from the TF
relation, as modified by mergers.

The TF and FJ relations relate the specific kinetic energy required to
support a galaxy to the attractive potential well created by the mass
of the galaxy.  Using the DEIMOS multi-object spectrograph at Keck
Observatory, \citet{Kassin} observed the kinematics and masses of
emission line galaxies in a redshift range from 0.1 to 1.2 from the
All Wavelength Extended Groth Strip International Survey (AEGIS),
incorporating morphologically normal and disturbed galaxies.  They
find that the disturbed galaxies do not lie on the standard TF
relation which only considers global rotational support.  Many of the
disturbed galaxies have rotation velocities that are well below the
normal values.  This morphological dependence of the TF scatter is
well known.  Studies of local galaxies have found larger scatter for
close pairs of galaxies with kinematic disorders \citep{bart} and when
peculiar galaxies are added \citep{kann}.  At $z\sim 0.5$,
\citet{flor} find larger TF scatter for galaxies with ``complex
kinematics,'' which was confirmed by the larger sample and more
detailed analysis of \citet{puec}.  This correlation of scatter with
morphology is not surprising, as one might expect kinematic scaling
relations based on pure rotational support to fail for cases where
the systems are not in globally ordered motion.  Galaxies that are
merging, disturbed, or rapidly evolving may exhibit disordered
kinematics, so that considering only rotation will not capture all
the kinematic support that balances the gravitational attraction of
the potential.

Interestingly, when \citet{Weiner06a} and \citet{Kassin} incorporated
the random internal motions of galaxies, as measured by velocity
dispersion, along with rotational velocity into a new parameter,
$S_{0.5}=\sqrt{0.5 V_{\rm rot}^2 + \sigma^2}$, then all galaxies
regardless of morphology fell onto a single relation between ${\rm
  S_{0.5}}$ and stellar mass.  This new parameter allowed for all
galaxies to be incorporated in a measurement of the TF relation,
instead of requiring the TF to be restricted to a sub-sample of
galaxies selected to be morphologically and/or kinematically
well-ordered.  In addition, this new TF relation is consistent with
the FJ relation for elliptical galaxies from \citet{gall}.  It is
perhaps surprising that a set of non-homologous systems fall on the same
kinematic relation, suggesting that the relation contains information
on the kinematic evolution of galaxies.

Given the complexity of effects at work, the constraints that produce
this tight kinematic relation are not clear.  The velocity dispersion
as measured in gas-phase kinematics by \citet{Weiner06a} and
\citet{Kassin} appears to provide kinematic support similar to a
thermalized velocity dispersion, but it probably represents spatially
unresolved velocity gradients rather than truly randomized motions.
Observational effects, such as blurring, can transform chaotic motions
or spatially unresolved rotation into apparent dispersion.  The
$S_{0.5}$ quantity is formulated to measure a total specific kinetic
energy; although these systems are non-equilibrium, it is likely that
the virial theorem remains a good approximation for the balance
between kinematics and gravitational potential.  In order to
investigate these possibilities and to explore the origin of the ${\rm
  S_{0.5}}$-stellar mass relation, we analyze a large suite of
hydrodynamical galaxy merger simulations.  Mock observations of this
simulation set provide kinematic data on isolated disk galaxies,
disturbed galaxies, merging/overlapping galaxies, and elliptical
merger remnants.  The simulations allow for a comparison with
observational results as well as a parallel analysis of intrinsic
kinematics and observational effects.

\section{Methods}

\subsection{Galaxy Merger Simulations}
We exploit a large suite of binary galaxy merger simulations
originally used to study feedback and star-formation in galaxy
mergers.  It provides a rich source of kinematic data for galaxies
with a wide range of morphologies.  For this study use a selection of
50 merger simulations from this suite.  For full details on the merger
simulations we refer the reader to previous studies \citep{Cox04,
  thesis, Cox05, minors}, but we include a brief description here for
completeness.

The numerical simulations performed in this work use the N-Body/SPH
code GADGET \citep{SpGad}.  Hydrodynamics are included via the
Lagrangian technique of smoothed particle hydrodynamics (SPH).  We use
the ``conservative entropy'' version of SPH \citep{SpEnt}.  Gas is
assumed, for simplicity, to be a primordial plasma that can
radiatively cool via atomic and free-free emission.

All of the numerical simulations presented here include star
formation.  Stars are formed in regions of gas that are above a
critical density for star formation at a rate proportional to the
local gas density and inversely proportional to the local dynamical
time-scale.  The efficiency of star formation is fixed by requiring
star formation to follow the observed correlation between gas surface
densities and star-formation rate \citep{Kenn98}.

We also include a simple prescription to simulate the effects of
feedback from massive stars.  This feedback acts to pressurize the
interstellar medium and regulates the conversion of gas to stars.
Details of this model and the parameter choices can be found in
\citet{Cox05}.  Specifically, the simulations studied in this paper
used the $n2med$ parameter set.  Under these assumptions the gas
pressure increases as the density squared; i.e, star-forming gas has a
``stiff'' equation of state.  

The simulations adopt a gravitational softening length of $h=400$~pc
for the dark matter particles and 100~pc for the stellar and gas
particles.  We remind the reader that, in GADGET, forces between
neighboring particles become non-Newtonian for separations $<2.3$
times the gravitational softening length.

The disk galaxy models and orbits are cosmologically motivated, but
the simulations are not cosmological since the two galaxies are
isolated.  All mergers are binary mergers of disk galaxies.  Rather
than being produced by previous mergers, the progenitor disk galaxies
are constructed in equilibrium and contain dark matter, an exponential
stellar disk, an extended exponential gas disk, and some contain a
dense central bulge.  The dark matter halos are assumed to have an NFW
profile, whereas the stellar bulges have a three-dimensional
exponential profile.  Bulge to disk stellar mass ratios vary from 0.02
to 0.25. The portion of the suite used in this study contains two
types of progenitor galaxy models:
\begin{enumerate}
\item ``Sbc'' galaxies, which are modeled after local Sbc-type
  spirals, with a small bulge and an initial gas mass roughly equal to
  the initial stellar mass.
\item ``G'' galaxies, which span a range of mass, bulge fraction, and
  gas fraction. Their properties are taken from statistical samples of
  local late type galaxies, including the Sloan Digital Sky Survey
  \citep{York00}.
\end{enumerate}

The Sbc mergers are all major mergers of identical progenitors with a
stellar disk of $3.92 \times 10^{10} \msun$, a gas disk of $5.36
\times 10^{10} \msun$, a stellar bulge of $1.00 \times 10^{10} \msun$,
and a dark matter halo of $81.2 \times 10^{10} \msun$.  The Sbc series
contains the widest variation of initial merger orbits and
orientations, with 17 different orbits run. This includes orbits with
a wide range of pericentric distances, eccentricities between 0.6 and
1.0, and a variety of progenitor orientations, including
prograde-prograde, prograde-retrograde, retrograde-retrograde, and
some cases inbetween. The G series mergers include both major mergers
between identical galaxies and mergers of galaxies with mass ratios
between 1:1 and 1:50.  For more detail on these models see
\citet{thesis}, \citet{minors}, and \citet{remnants}.

\subsection{Mock Observations of the Simulations}

In order to make comparisons with the observational results, we must
first mock `observe' the galaxies in the same manner as
\citet{Kassin}.  They used an algorithm called ROTCURVE
\citep{Weiner06a}.  This algorithm was designed to obtain kinematic
measurements for as many galaxies as possible by allowing model
seeing-compensated rotation curves to be fit using only the 2-d
spectra.  ROTCURVE creates an intrinsic model for the emission
intensity, velocity, and dispersion, blurs this model to simulate
seeing, and fits the model to the data.  The intrinsic model used
is:
\begin{equation}
I(x,v)=G(x)\exp(-\frac{(v- V(x))^2}{2\sigma_{\rm 2d}^2}),
\end{equation}
\begin{equation}
G(x)= \frac{I_{\rm tot}}{\sqrt{2\pi}r_{\rm i}}\exp(-\frac{(x-x_0)^2}{2r_{\rm i}^2}),\\
\end{equation}
\begin{equation}
V(x)= V_{\rm rot}\frac{2}{\pi}\arctan(x/r_{\rm v}),
\end{equation}
where $G(x)$ is an assumed Gaussian light distribution along the slit,
$V(x)$ is the rotation curve, with asymptotic velocity $V_{\rm rot}$
and knee radius $r_{\rm v}$, and $\sigma_{\rm 2d}$ is the velocity
dispersion, which is assumed to be constant along the slit.  In order
to fit the spectra, $I(x,v)$ is blurred using a 1-d Gaussian in the
spatial direction.  After blurring, moments are taken of the blurred
model in the velocity direction in order get values of average
velocity and dispersion for each bin within the slit.  These values
are then compared against the observed rotation curve by computing a
$\chi^2$ value.  $\chi^2$ is then minimized via adjustments of $V_{\rm
  rot}$ and $\sigma_{\rm 2d}$.  The other potentially adjustable
parameter, $r_{\rm v}$, is held fixed at 0.2 arcsecs. Observational
results do not change significantly if $r_{\rm v}$ is varied within a
reasonable range of 0.1 to 0.3 arcsec.

In order to analyze the simulations, we use an analogous algorithm.
Each simulation has snapshots every 40-100 Myr, and each of these
snapshots is viewed from 11 different evenly distributed angles. The
simulations are assumed to be at three different redshifts
($z\sim0.03$, $z\sim0.3$, and $z\sim1.0$), which approximately cover
the redshift range in Kassin et al.  We include the lowest redshift
bin to examine the relation with very little blurring.  This results
in a sample of roughly 150,000 mock observations.  Physical distances
are converted to arcsecs given the assumed redshift of the
observation, using cosmological parameters $h=0.7$, $\Omega_m=0.3$,
and $\Omega_\Lambda=0.7$.  The effects of seeing are taken into
account by smearing particle positions using a Gaussian with
$\sigma=0.3~{\rm arcsecs}$.  If the projected separation of the two
galaxies is less than 0.7, arcsecs then the two galaxies are treated
as one object.  This allows us to include merging and overlapping
galaxies in the sample.  We orient the mock slit by fitting an ellipse
to the simulated stellar density profile.  All mock observations use a
slit width of 1.0 arcsec as in the observations. The mock slit length
varies with redshift beginning with a length of 4.0 arcsecs at
$z\sim1$ and scaling with physical galaxy size, as in the
observations.

Within the slit, stellar particles - either all stars or just new
stars (see Section \ref{ssec:newstars}) - are separated into 0.1
arcsec bins, mimicking the pixel width in the observations.  Rotation
curves are calculated using the average velocities and dispersions of
the particles in each bin.  We fit $G(x)$ to the bin stellar masses
and calculate $I(x,v)$ and $V(x)$ as above.  Unlike \citet{Weiner06a}
and \citet{Kassin}, we do not hold $r_{\rm v}$ fixed, but instead
assume a value of 0.2 arcsecs at $z \sim 1$ and scale that value with
the physical size of the galaxy.  If we do not scale $r_{\rm v}$ and
the slit length with redshift, we get spurious systematic offsets in
$V_{\rm rot}$ when comparing the lowest redshift bin with the two
higher redshift bins.  Inclination is also calculated, using the thin
disk approximation as in the observational work, and is used to
correct $V_{\rm rot}$ to edge-on. We remove all galaxies from the
sample with inclinations less than 50 degrees, as we find that the
inclination fits become unreliable below that value.  Similarly,
Kassin et al. remove the galaxies with inclinations less than 30
degrees.  The observational sample is also cut for nearly edge-on
disks (inclination greater than 70 degrees) because of dust effects on
stellar mass estimates.  Since the simulations do not suffer from
this same limitation, we do not cut at large inclinations.

\section{Comparison of the Simulations with Observations}

\subsection{Evolution of a Single Galaxy Merger}
\label{ssec:single}
To illustrate how the kinematics of galaxies change during a merger
event, we follow the time evolution of a single galaxy merger.  Even
though this is only a specific example case from the set of
simulations studied, it is typical of the mock observations of the
other simulations, as evidenced by the results of Section
\ref{sec:ensemble}. The simulation used for this study is a merger of
two identical Sbc galaxies.  Initially, the galaxies are set on a
parabolic prograde-prograde orbit with a pericentric distance of 11.0
kpc.  We depict the evolution of kinematic quantities during the
merger viewed from a single angle (Figure 1), and
follow the galaxies' evolution on the kinematic relations (Figure
2, numbers).  The merger is assumed to be at a redshift of
one.


We start off at time=0 Gyr with the progenitors, which are designed to
lie on the observed TF ridge-line. The progenitor is demarcated as `1'
in Figures 1 and 2.  We follow one of the
galaxies until the projected distance between the galaxies becomes
less than 0.7 arcsecs after which we follow the merger as a single
system.  The merging galaxies have two encounters and merge during the
second encounter.  The first encounter occurs at 0.6 Gyr after the
start of the simulation.  This encounter results in a significant
increase in the velocity dispersion by $\sim60~ {\rm km~s^{-1}}$ and a
sharp decrease in the rotation velocity by $\sim55~ {\rm km~s^{-1}}$.
During this encounter, the galaxy appears morphologically disturbed,
especially in its outer disk.  Shortly following the encounter, at 0.7
Gyr, the galaxy is a slightly low rotation velocity outlier in the TF
(`2' in Figures 1 and 2).  Between the
first and second/final encounters, the rotation velocity of the galaxy
gradually increases by $\sim20~ {\rm km~ s^{-1}}$ as tidal debris that
was removed from the disk settles back.  The final coalescence occurs
at 1.75 Gyr after the start of the simulation (`3' in Figures
1 and 2). At coalescence, the galaxy's
rotation velocity drops abruptly to a meager 20 ${\rm km~s^{-1}}$.
The mock observational slit now contains both the first progenitor and
the remains of the second progenitor, which has been severely
disrupted.  Projection results in the overlapping of the two
progenitors within the slit and pollutes the rotation curve, causing
the dramatic decrease in rotational velocity.  Consequently, the
system lies furthest from the TF law at this time (`3' in Figures
1 and 2).  At the final coalescence, we
find an increase in the velocity dispersion by $\sim 100~{\rm
  km~s^{-1}}$.  From this point on, the kinematics are dominated by
dispersion rather than rotation. The final stage (`4' in Figures
1 and 2) depicts the merger remnant, which
is a rotating elliptical galaxy.  In contrast to the other kinematic
quantities, ${\rm S_{0.5}}$ remains relatively stable throughout the
merger process, increasing primarily when the two progenitors combine
to create a more massive system.

The tendency for rotation velocity to decrease as velocity dispersion
increases during galaxy interactions and mergers provides a simple
mechanism to explain the observed ${\rm S_{0.5}}$-stellar mass
relation and large scatter in the conventional TF.  We note, however,
that observed ${\rm S_{0.5}}$ is not strictly conserved, even when
mass is constant.  During a close encounter all of the `lost' rotation
is not immediately converted into velocity dispersion.  Rather, the
tendency during such an encounter is for the apparent rotation
velocity to briefly drop to very low values while the velocity
dispersion increases only modestly.  The effects of this lack of
strict conservation of ${\rm S_{0.5}}$ fall within the scatter of the
${\rm S_{0.5}}$-stellar mass relation.


\subsection{Kinematic Relations for an Ensemble of Simulated Galaxy Mergers}
\label{sec:ensemble}
In order to further test this explanation for the observed kinematic
relations, we plot rotation velocity and ${\rm S_{0.5}}$ versus
stellar mass for 500 randomly chosen snapshots from set of 50 merger
simulations and compare to similar plots for 544 galaxies from
\citet{Kassin} (Figure 2). Each of these snapshots is
viewed from an angle chosen at random from the 11 different evenly
distributed viewing angles. These 500 snapshots are analyzed assuming
$z\sim0.03$, $z\sim0.3$, and $z\sim1.0$.  We find that at all
redshifts the simulated TF has a significant scatter to low $V_{\rm
  rot}$ (spanning $\sim1.5$ dex), similar to the observations.
Furthermore, this scatter is correlated with close encounters.
Sixty-four percent of the cases with scatter greater than 0.5 dex are
encounters/mergers, whereas encounters/mergers are only 30\% of the
entire sample.  The primary difference between the simulated and
observed results is in the correlation between scatter and redshift.
For the observations, scatter in the TF significantly increases at
high redshift, whereas for the simulations, the scatter in the TF does
not correlate strongly with redshift.  This suggests that the observed
evolution in scatter of the TF represents a real change in kinematics
with redshift as opposed to an observational blurring effect.

If we incorporate random internal motions using ${\rm S_{0.5}}$, then
we find a single, relatively tight relation with stellar mass.
Comparing different redshift bins we see that the slope fitted to the
simulated galaxies stays within a range of $\sim 2.8-2.9$, whereas the
observed slopes are typically slightly steeper but range between $\sim
2.4-3.3$.  There is no significant trend in slope with redshift in
either the observations or simulations.  In the two higher redshift
bins the scatter in ${\rm S_{0.5}}$ for the simulations is 0.08-0.09
dex.  The intrinsic scatter in the observed relation ranges between
0.08 and 0.12 dex and also has no systematic trend with redshift.
The lowest redshift bin for the simulated results does show more
scatter (0.16 dex), but this is likely due to difficulty in fitting
unblurred rotation curves of disturbed objects, which often do not
look like the idealized arctan rotation curve.  This is also outside
of the range of the observations discussed in \citet{Kassin}.

While a detailed quantitative comparison of the samples is not
warranted because the galaxies in the simulation suite are not a
statistically representative sample of the types of galaxies that one
would expect to find in the real Universe, the agreement with the
observed relations demonstrates that galaxy mergers and interactions
are a mechanism that can create large scatter in the TF concurrent
with low scatter in the ${\rm S_{0.5}}$-stellar mass relation.

Furthermore, the simulated results show that the merging process moves
galaxies up the ${\rm S_{0.5}}$-stellar mass relation toward higher
values of ${\rm S_{0.5}}$, and therefore that the relation ties
together progenitors (spirals), merging galaxies, and merger remnants
(ellipticals).  The robustness of this relation has two causes.
First, throughout the merger process there is a tendency for rotation
and velocity dispersion (or, random internal motions) to be
anti-correlated.  It has been known for a long time that galaxy
mergers are an effective mechanism for converting ordered rotational
support into random pressure support \citep{TT72}. As noted above, it
is surprising that a kinematic relation would hold for even merging
cases.  This results from a subtly different cause.  Much of the
apparent rotation velocity may be lost in an encounter when the two
galaxies overlap in projection.  This is because the two galaxies'
orbital velocities and internal velocities rarely add coherently when
randomly overlapped.  However, velocity dispersions are not so easily
lost.  In fact, overlapping two galaxies with interfering rotation
curves will increase velocity dispersion.  Thus, when such an overlap
occurs, apparent rotation typically decreases, whereas velocity
dispersion typically increases.

\section{Observational Effects on the ${\rm S_{0.5}}$-Stellar Mass Relation.}

From the observations alone, it is unclear the extent to which the
${\rm S_{0.5}}$-stellar mass relation is a result of observational
effects or is telling us something about the intrinsic properties of
the galaxies.  In this section, we explore the various observational
effects on the kinematic measurements of the simulations and determine
whether they can contribute to the observed relation.

\subsection{Blurring}

As galaxies are observed at higher redshifts, their angular extent
becomes smaller.  Consequently, the typical $\sim 0.7''$ seeing
produced by the atmosphere in the Kassin et al. observations
effectively moves the stars from one spectral bin into other nearby
bins within the slit.  This reduces the apparent rotation velocity and
produces an artificial velocity dispersion \citep{Weiner06a}.  The
fitting algorithm used by ROTCURVE takes this into account by fitting
a blurred model and extrapolating back to an intrinsic model.
However, an interesting question is how successful this procedure is,
and whether or not blurring conserves ${\rm S_{0.5}}$.  If ${\rm
  S_{0.5}}$ is conserved by blurring, then it is possible that
blurring is partially responsible for the scatter in the TF and the
tight ${\rm S_{0.5}}$ relation.

In order to explore the effects of blurring, we take the fiducial
merger simulation from \S \ref{ssec:single} (the Sbc) and analyze
the kinematics after blurring the particles using gaussians with
$\sigma$'s of 0.0'', 0.1'', 0.3'', 0.6'', 0.8'', and 1.6'', assuming
the galaxy is at a redshift $z\sim 1$ (Figure \ref{fig:blur}).  We use
the four example snapshots from Figure 1: the
undisturbed disk, disturbed disk, merger, and merger remnant.  The
effective blurring within the AEGIS sample is all between 0.1'' and
0.3'', given seeing of 0.7'' (defined as full-width at half maximum).
There is little difference between the unblurred kinematics and 0.1''
kinematics for all snapshots shown. However, for both the disks and
the remnant there is a decrease in $V_{\rm rot}$ and increase in $\sigma$
by the time we have reached 0.3''.  However, for all cases, the
decrease in $V_{\rm rot} \leq 30\%$.  Also, for all cases, as the blurring
becomes much greater than $R_{\rm 50}$, the radius that contains
$50\%$ of the stellar mass, the ROTCURVE algorithm breaks down and
$V_{\rm rot}$ drops to very low values while $\sigma$ rises.

For the undisturbed disk, disturbed disk, and merger, ${\rm S_{0.5}}$
is approximately conserved by blurring as $V_{\rm rot}$ and $\sigma$
adjust in lockstep.  However, for the remnant, by 0.3'' blurring, ${\rm
  S_{0.5}}$ has already increased by $\sim 50\%$.  This results from a
steeply-peaked central velocity dispersion.  For zero blurring, the
assumption of a constant $\sigma$ underestimates the kinetic energy
contribution from random motion because the central peak is small with
respect to the size of the slit.  As this peak gets blurred, the fit
improves.

Therefore, it is true that as blurring becomes very large
($>>R_{50}$), the ROTCURVE fitting procedure will break down and the
rotational component of the kinematics will transform into dispersion.
Also, for three of the four cases shown, ${\rm S_{0.5}}$ is conserved
throughout this process.  Therefore it is true that blurring could
cause some additional spread in the TF and that using ${\rm S_{0.5}}$
rather than rotation velocity would help to reduce this spread.
However, for the range of blurring and galaxy sizes in the
observational study, the ROTCURVE model performs relatively well, and
certainly does not produce large enough errors to low $V_{\rm rot}$ to
account for the observed scatter.  This is also confirmed by the
apparent lack of systematic changes in $V_{\rm rot}$ with redshift in
Figure 2.  Thus blurring cannot be the main reason for the
observational findings.

\subsection{Tracking of Emission Lines}
\label{ssec:newstars}
A potential concern with our analysis is that the AEGIS observational
results rely on OII 3727 emission line spectra from hot gas, whereas
the kinematic analysis of the simulations described above uses 
the star particles.  Specifically, the line emission is known to be
correlated with areas of new star formation.  Thus, we repeat the
analysis from \S \ref{sec:ensemble} using only new stars that form in
the simulation and assuming $z=1$ (Figure \ref{fig:ns}).  This gives
us many fewer particles to work with, especially in the early stages
of the merger when few new stars have formed.  In order to provide
sufficient statistics within the slit, we restrict the analysis to
snapshots that have greater than 500 new star particles.  This removes
some of the sample from early on in the simulations, but will provide
a check for whether or not limiting the analysis to new stars affects
the results.

The TF and ${\rm S_{0.5}}$ relations are qualitatively quite similar
to those calculated using all star particles, showing large scatter
for TF and relatively small scatter for ${\rm S_{0.5}}$.  This
suggests that the observational difference between tracing kinematics
using hot gas and stars does not have a significant effect on the
results.  There is some additional scatter in the new star ${\rm
  S_{0.5}}$ relation (0.14 dex as compared to 0.08 dex).  However, one
must also remember that dust, which is not taken into account in the
simulations, creates a counteracting effect.  Within the simulations,
analyzing only new stars will emphasize the central portions of the
galaxies (where most stars form), however, these also tend to be the
portions of the galaxy most enshrouded by dust.  Thus, the real
difference between observing emission lines (gas) or absorption lines
(stars) is likely to be less.  In order to truly pin down the effect
of dust, one would have to create artificial spectral kinematics by
fully modeling the radiative transfer through the gas and dust.  This
will soon be possible using a newer version of SUNRISE (Jonsson, 2006
and in preparation).


\section{Intrinsic Kinematic Quantities}

We have explored some of the observational effects on the kinematic
relations, however, we can further utilize the simulations to examine
intrinsic kinematic quantities in order to better understand the
origin of the ${\rm S_{\rm 0.5}}$-stellar mass relation.  If one
assumes a tracer population within an isothermal sphere at a distance
$r>>r_{\rm core}$, where $r_{\rm core}$ is the core radius of the
isothermal sphere, and then allows that population to assume complete
rotational support or complete support from random motion then one
finds that $V_{\rm circ}\sim \sqrt{2}\sigma$, where $V_{\rm circ}$ is
the velocity with pure circular support and $\sigma$ is the dispersion
for pure random support \citep{Weiner06a, Kassin}.  This was the
motivation for the multiplying factor of 1/2 in front of $V_{\rm rot}$ in
${\rm S_{0.5}}$.  One might expect that a smooth transition between
rotational and random support, while conserving the density profile,
would result in conservation of ${\rm S_{0.5}}$.  Therefore, an
important question is whether this idealized picture is correct and
${\rm S_{0.5}}$ really does trace the mass distribution.  If so, this
would provide a means of estimating total masses of galaxies including
the contribution from dark matter.

\subsection{Kinematics as a Function of Radius}

We begin our analysis of intrinsic galaxy properties by examining
kinematics as a function of radius.  The ${\rm S_{0.5}}$ parameter is
motivated by results obtained from the Jeans Equations for an
isothermal sphere.  However, our systems are not isothermal spheres,
so we take a step back to see where the assumptions could be wrong.
For steady state spherically symmetric systems with no rotation, the
following relation holds (Binney and Tremaine, 1987, eq. (4-55)):
\begin{equation}
\label{eqn:jeans}
V_{\rm circ}^2= \frac{GM(r)}{r} = -\overline{\sigma_{\rm r}^2}\left( \frac{d~{\rm ln~}\nu}{d~{\rm ln~}r} + \frac{d~{\rm ln~}\overline{\sigma_{\rm r}^2}}{d~{\rm ln}~r} + 2\beta\right),
\end{equation}
where $\sigma_{\rm r}$ is the radial velocity dispersion, $\nu$ is the mass
density, $\beta \equiv 1 - \sigma_\theta^2/\sigma_{\phi}^2$ is the
velocity anisotropy parameter, and $\sigma_\theta$ and $\sigma_\phi$
are the velocity dispersions in the $\theta$ and $\phi$ directions.
At a given radius, this can be represented as the simpler
\begin{equation}
V_{\rm circ}^2=k \sigma_{\rm r}^2.
\end{equation}

For the singular isothermal sphere, the second two terms in the
parentheses are zero and we are left only with
$V_{\rm circ}^2=-\overline{\sigma_{\rm r}^2}(d~{\rm ln}~\nu/d~{\rm ln}~r)$.  For
an isothermal sphere with a core radius $r_{\rm core}$ and $r>>r_{\rm
  core}$ the density term $(d~{\rm ln}~\nu)/(d~{\rm ln}~r) \sim -2$, giving
us the constant, $k\sim2$, in the ${\rm S_{0.5}}$ relation.  For our
simulation set, and for real galaxies, we are neither guaranteed that
the density term will be $\sim -2$ nor that the other two terms will
vanish.  Thus we look in detail at the values these terms are likely
to have.

The velocity dispersion is generally a weak function of radius.  Thus
the dispersion term, $d({\rm ln~}\overline{\sigma_{\rm r}^2})/d({\rm
  ln}~r)$ is likely to be relatively small.  The drop in dispersion
with radius is quite small in the progenitors and is largest for the
remnants, where significant dissipation and star formation may have
steepened the profiles.  As shown in \citet{DekelDM} the dispersion in
the remnants is relatively well represented by $\sigma \propto
r^{-0.2}$.  Thus the contribution of the dispersion term in equation
\ref{eqn:jeans} is likely to be $-0.2$ or smaller, which is quite
small compared to the presumed value of the density term.  Because of
tidal disturbance and dissipation, the stellar orbits in the outer
portions of the merger remnants do exhibit relatively large
anisotropies with $\beta \sim 0.2-0.5$ \citep{DekelDM}.  Furthermore
this term is multiplied by a factor of two bringing the overall
contribution to $\sim 0.4-1.0$.  For progenitors, which typically have
low anisotropies, one would expect the contributions from these terms
to roughly cancel, whereas for remnants there would be a net
contribution to $k$ of $\sim 0.2-0.8$.

The largest of the three terms is the density term.  This term has a
value of roughly two for an isothermal sphere and three or four for
the outer portion of an NFW or a Hernquist profile.  Thus we very well
might expect a steeper density relation, $\rho \propto r^{-a}$, and
consequently higher value of $k$ than is found for the isothermal
case.  For each simulation we fit a value of $a$ to the total mass
density curve in the vicinity of $R_{\rm 50}$, the radius that contains
50\% of the stellar mass.  For the simulations, the value of $a$ at
$R_{\rm 50}$ ranges between $\sim 1.5-3.5$, with a peak around
negative two (see Figure \ref{fig:a}).  The slope is typically steeper for
massive progenitors and for merger remnants.  In summary, for
progenitors, the density term is likely to be $\sim 2$, while the
other terms will be small.  For remnants, the density term will often
be closer to $3$, but the anisotropy term is likely to counteract
much of this effect.  So for all cases, $k\sim2$ is a reasonable value,
but the variability in the density profiles and kinematics is sure to
produce variation around this value.

For our exploration of intrinsic galaxy quantities, we adopt a
quantity, $S \equiv \sqrt{V_{\rm rot}^2 + 2.0\sigma^2}$, which is just
$\sqrt{2}{\rm S_{\rm 0.5}}$ but with the benefit that it approaches
$V_{\rm rot}$ in the low-$\sigma$ limit.  This provides a quantity
that can be directly compared with the velocity due to all mass,
$V_{\rm circ}$.  The constant $k$ is a function of radius, and
therefore it is useful to determine the range in radius for which
$k\sim 2$, or alternatively $S\sim V_{\rm circ}$.  In order to examine
the value of $k$ as a function of radius, we stack all of the
simulated galaxies, normalizing radii to $R_{\rm 50}$.  For this
analysis, we separate the particles into bins of width equal to $0.2
R_{\rm 50}$.  For each bin, we determine an average velocity and
velocity dispersion for the stellar particles.  This provides us with
intrinsic quantities analogous to $V_{\rm rot}$ and $\sigma$.  These
values are used to calculate $S_{\rm intrinsic}$.  $V_{\rm circ}
\equiv \sqrt{G M(<r)/r}$ is calculated for each bin by totaling the
masses of all particles with a radius less than the bin, including
baryons and dark matter.  In Figure \ref{fig:ratio}, we plot the ratio
of $S_{\rm intrinsic}$ and $V_{\rm circ}$ as a function of radius
averaged over the velocity profiles of every snapshot within the
simulation suite. One can see that at small radii there is no
one-to-one correspondence between $S_{\rm intrinsic}$ and $V_{\rm
  circ}$ as the ratio ($S_{\rm intrinsic}/V_{\rm circ}$) becomes
closer to 2 or 3 on average with a large scatter between simulated
profiles.  However, at radii equal to or larger than $R_{\rm 50}$, the
ratio is quite close to unity and the scatter is about 30\% in either
direction.  Therefore, at radii larger than $R_{\rm 50}$, $S_{\rm
  intrinsic}$ is a good indicator of total enclosed mass.  For
simplicity, we choose a radius of $R_{\rm 50}$ at which to draw a
value of $S_{\rm intrinsic}$ for comparison across the entire sample.

\subsection{Correlation between $S_{\rm intrinsic}$ and $V_{\rm circ}$}

Now that we have demonstrated that $S_{\rm intrinsic}$ is roughly
comparable to $V_{\rm circ}$ at $R_{\rm 50}$, it is illustrative to
plot the relation between $S_{\rm intrinsic}$ and $V_{\rm circ}$ for
the sample.  We take these values at $R_{\rm 50}$ and plot them for
500 randomly chosen snapshots in Figure \ref{fig:intr}.  If the two
galaxies were separated by less than the sum of their stellar
half-mass radii then they were treated as one.  The $x=y$ line
represents equality between the two values.  One can see that $S_{\rm
  intrinsic}$ is a good indicator of enclosed mass.  The scatter in
the relation increases with mass.  This is the result of the
distribution of galaxy types in the simulations.  There are
considerably more simulations run for the larger galaxies, and these
simulations have a wide variety of orbits that can introduce
significant spread in remnant properties \citep{remnants}.  Therefore
our simulation set will automatically produce more spread at higher
masses.  However, even for the larger mass cases, whose spread is
likely more representative of the real universe than the low mass
cases, $\log(S_{\rm intrinsic}) \sim \log(V_{\rm circ})$ to typically
within 0.15 dex.

\subsection{Correlation between $S_{\rm observed}$ and Intrinsic $V_{\rm circ}$}

A final question of interest is whether or not the observed quantity,
$S$, can be used to estimate total galaxy mass, including the
contribution from dark matter.  To address this question, we plot the
relation between $S_{\rm observed}$ and the {\it intrinsic} $V_{\rm
  circ}$ in Figure \ref{fig:vcirc}.  Again, we have moved the factor
of 2 in the ${\rm S_{0.5}}$ quantity over to the $\sigma$ in order to
produce a correspondence between $S_{\rm observed}$ and $V_{\rm
  circ}$.  Here we calculated a projected $R_{50}$ using an elliptical
aperture and measure a three-dimensional $V_{\rm circ}$ within that
radius.  The $V_{\rm rot}$ values are taken as the value of the fitted
arctan curve at $R_{\rm 50}$ rather than the asymptotic value.  The
observations are assumed to be at a redshift of one.  As can be seen
in Figure \ref{fig:vcirc}, there is a rough correspondence between
$S_{\rm observed}$ and $V_{\rm circ}$.  Therefore, we take this result
as a confirmation that $S_{\rm observed}$ can be used to track central
galaxy mass, including contributions from dark matter.  There is a
tendency for $S_{\rm observed}$ to be a bit lower than $V_{\rm
  circ}$. This is likely the result of projection effects and
counter-rotation, both of which would serve to reduce $S_{\rm
  observed}$ compared with $S_{\rm intrinsic}$.  The outliers to low
$S_{\rm observed}$ are typically overlapping cases where the
kinematics are poorly represented by an arctan curve.  The fact that
the intrinsic relation is significantly better than the observed one
suggests that it may be possible to improve the observational
measurement of $S$.  The assumptions of the arctan rotation curve and
flat velocity dispersion often fail for the simulated galaxies,
particularly when two galaxies overlap.  Future work should examine
possible improvements to the observational algorithm.

\section{Discussion}

The observational results from \citet{Kassin} were interesting for
several reasons.  They demonstrated that a new kinematic quantity,
${\rm S_{0.5}}$, was capable of producing a relation that unified all
galaxy types, including disturbed and merging cases.  However, this
study left open several questions that cannot be answered from the
observations alone.  First, what sort of mechanisms could produce the
large scatter in the stellar mass TF while preserving the ${\rm
  S_{0.5}}$-stellar mass relation?  A second question is whether the
relation is telling us something deep about the observed galaxies, or
whether it is an artifact from the observations.  Here we used a suite
of galaxy merger simulations to address these questions.

We find that galaxy mergers and interactions are capable of
reproducing the observed relation.  Specifically, there are merger
stages during which the galaxies overlap and the rotation velocity
significantly drops.  This overlap results in increased dispersion
which when included in a kinematic relation helps to reduce the
scatter.  Additionally, within a given merger one can see that
rotational support is transformed into random pressure support.  This
transformation occurs in such a way that the scaling between
$\log({\rm S_{0.5}})$ and $\log(\rm M_{\rm star})$ is roughly conserved.
Thus, galaxy mergers move galaxies up along the ${\rm
  S_{0.5}}$-stellar mass relation, and the relation effectively
unifies progenitors, mergers, and remnants.

We examine the observational effect of blurring to see whether it
could be responsible for the observed additional dispersion and lack
of rotation.  We find that the ROTCURVE algorithm performs well for
the range of blurring found in the observations.  We also find that
${\rm S_{0.5}}$ is typically conserved by blurring.  Therefore it
could be responsible for some of the reduction in scatter, but the
effect is much too small to account for the small observed scatter in the 
${\rm S_{0.5}}$-stellar mass relation.  

Our examination of the kinematic scaling laws of the galaxies, and how
they change with radius suggests that the appropriate constant in the
${\rm S_{0.5}}$ parameter is $\sim 2$.  Additionally, we find that $S$
at $R_{\rm 50}$ is a reasonably good tracer of $V_{\rm circ}$ at
$R_{\rm 50}$.  Therefore, it can be used to infer the galaxy mass
within $R_{\rm 50}$, including contributions from dark matter.
However, the details of this relation warrant more study.
Specifically, the relation between $S_{\rm intrinsic}$ and $V_{\rm
  circ}$ is much better than that with $S_{\rm observed}$, suggesting
that it may be possible to improve the observational measurement of
$S$ and consequently of the total enclosed mass.  Additionally, it
would be interesting to compare the merger simulations with available
integral field unit observations of nearby, $z\sim0.6$, and $z\sim 2$
galaxies \citep[e.g.][respectively]{arri, yang, fors}.  It is also
likely that kinematics could be used as an indicator for merging
galaxies, and possibly merger stage, since mergers often produce TF
outliers and typically have significant asymmetry in their kinematics,
but the development of new kinematic indicators of merging is left for
future work.

\section{Acknowledgments}

MDC acknowledges support from NASA ATP grant NNX07AG94G, the UCSC
Physics Department, and a UCSC UARC-ARP grant.
JRP was supported by NASA ATP grant NNX07AG94G.
Simulations analyzed here were carried out at NERSC, NASA's Columbia 
supercomputer, and UCSC computer clusters.

\bibliographystyle{apj}
\bibliography{tj,matt,patriks,susan}

\begin{figure}
\begin{center}
\includegraphics[width=0.85\textwidth]{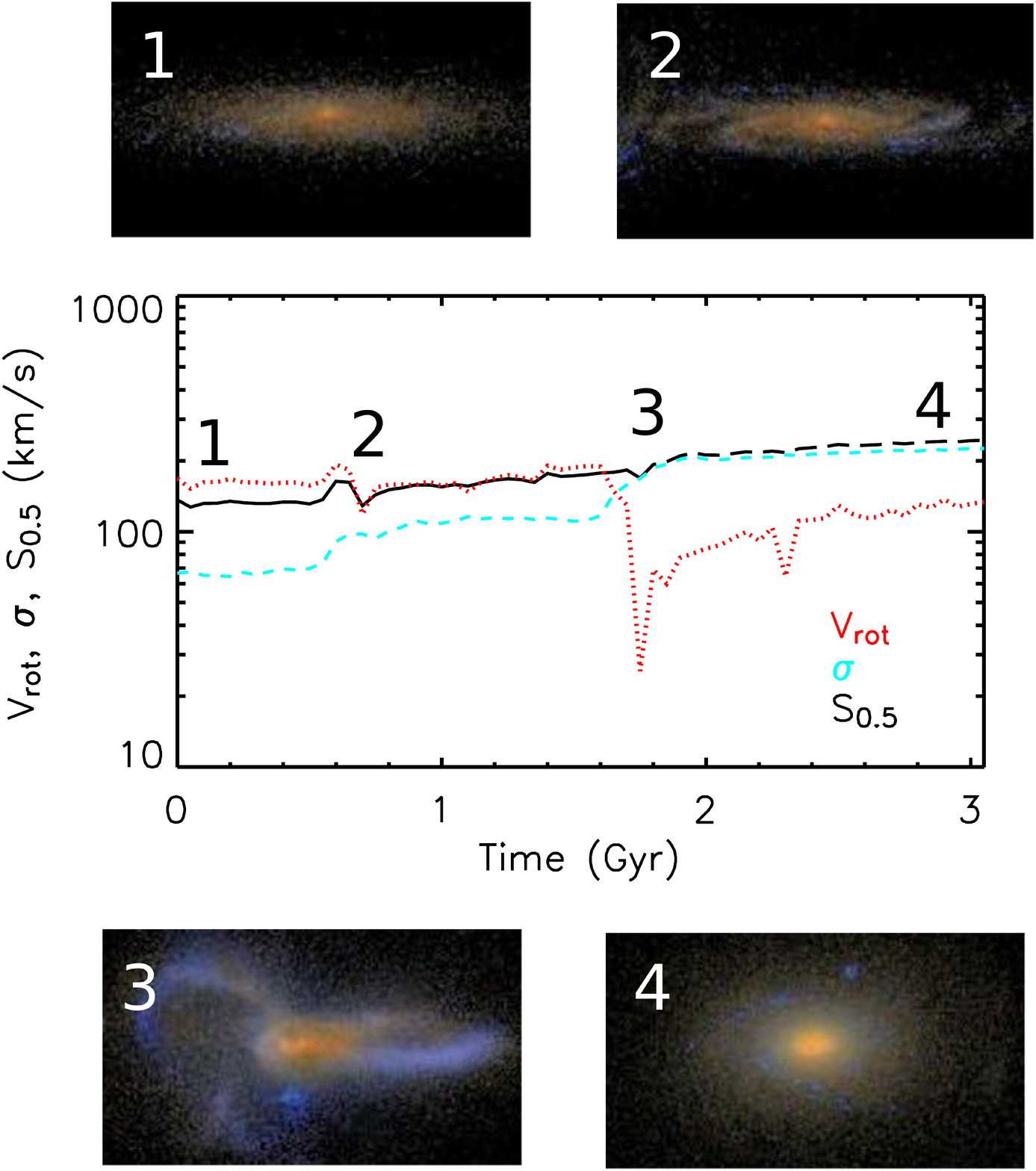}\\
\caption{}
\end{center}
\end{figure}
\begin{figure}[p]
\begin{center}
  \captcont[Time evolution of rotation velocity, velocity dispersion,
    and ${\rm S_{0.5}}$ during a single merger simulation]{Time
    evolution of rotation velocity (red dotted line), velocity
    dispersion (blue dashed line), and ${\rm S_{0.5}}$ (solid and
    long-dashed black lines) during a single merger simulation of two Sbc
    galaxies with stellar masses of $\sim5 \times 10^{10} \msun$
    initially on a parabolic orbit.  With each encounter, the rotation
    velocity decreases and the integrated velocity dispersion
    increases.  However, ${\rm S_{0.5}}$ remains relatively constant
    throughout the merger, increasing primarily when the two galaxies
    coalesce. Care is taken to mock observe the simulations just as in
    the actual observations, assuming $z=1$.  The solid portion of the
    ${\rm S_{0.5}}$ line denotes the snapshots where only a single
    progenitor is observed in the slit, and the long-dashed portion
    denotes snapshots where both progenitors are observed in the slit.
    The merger simulation uses the smoothed-particle hydrodynamics
    code GADGET \citep{SpGad} and includes gas, star formation, and
    stellar feedback.  The galaxy images are calculated from the
    simulation using the dust radiative transfer code SUNRISE
    \citep{Jonsson06} and are shown at four different times during the
    merger: 1) Before any encounters, 2) shortly after the initial
    encounter, 3) at the final coalescence, and 4) the remnant after
    the galaxies have coalesced.} 

\label{fig:kin_evo}
\end{center}
\end{figure}

\begin{figure}
\begin{center}
\includegraphics[width=0.85\textwidth]{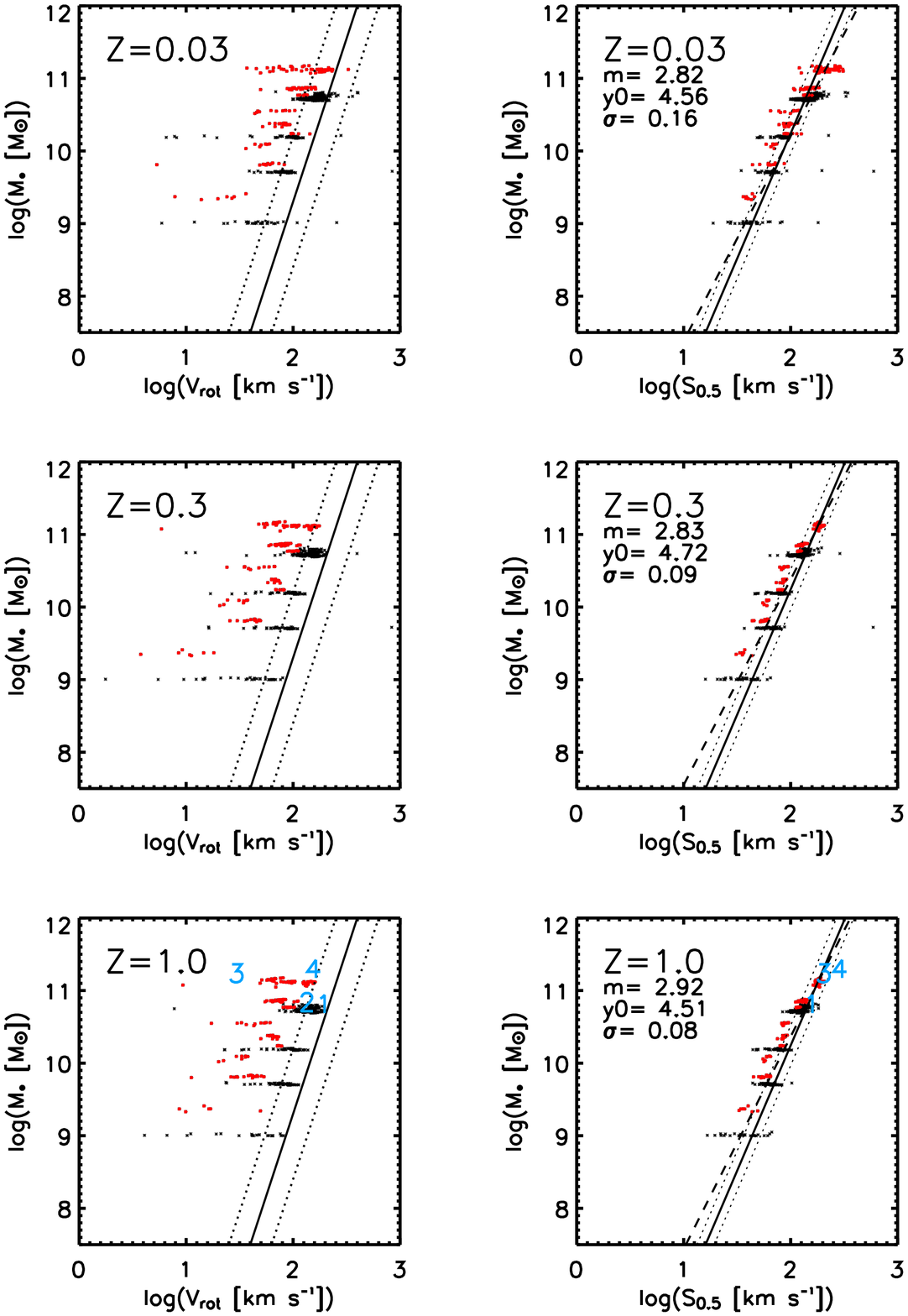}\\
\caption{}
\end{center}
\end{figure}
\begin{figure}[p]
\begin{center}
\captcont[The TF relation and ${\rm S_{0.5}}$-stellar mass
  relation for mock observations of simulated merging galaxies]{ The TF relation (left) and ${\rm S_{0.5}}$-stellar mass
  relation (right) for mock observations of simulated merging galaxies
  at redshifts of $z\sim0.03$ (top), $z\sim0.3$ (middle), and
  $z\sim1.0$ (bottom).  Each dot represents a mock observation of a
  single snapshot viewed from a single angle.  In order to compare
  with the observational results from \citet{Kassin}, 500 such images
  were chosen from the simulation set at random.  Red points are mock
  observations where both galaxies are present in the slit (i.e. close
  encounters and the merger remnant), and black points are mock
  observations of single simulated galaxies.  In the TF plots, the
  solid line is the high-redshift TF ridge line from
  \citet{Conselice05}, with the dotted lines representing the scatter.
  As in \citet{Kassin}, a significant number of the simulated galaxies
  scatter to low $V_{\rm rot}$.  In the observations, these galaxies have
  disordered or compact morphologies.  Similarly, in the simulations,
  the majority of the cases scattered to low $V_{\rm rot}$ are either
  undergoing or have recently undergone an encounter.  In the ${\rm
    S_{0.5}}$ plots, the solid line is the fit to the observed ${\rm
    S_{0.5}}$ relation at $z\sim1.0$, and the dotted lines depict
  scatter in the relation.  The dashed line is the best fit to the
  simulations.  Slopes ($m$), zero points ($y_0$), and scatters ($\sigma$)
  for each fit are listed ($y= y_0 + m x$).  Including velocity dispersion greatly
  reduces the scatter and brings the progenitors, disturbed galaxies,
  merging galaxies, and merger remnants onto a single kinematic
  relation.  The relation and scatter found for the simulated galaxies
  are comparable to the observed relation.  Numbers on the plots show
  the location of the various numbered merger stages from Figure
  1.  The bottom panel has no `2' because it overlaps
  `1'. }
\label{fig:tf}
\end{center}
\end{figure}

\begin{figure}
\begin{center}

\includegraphics[width=0.8\textwidth]{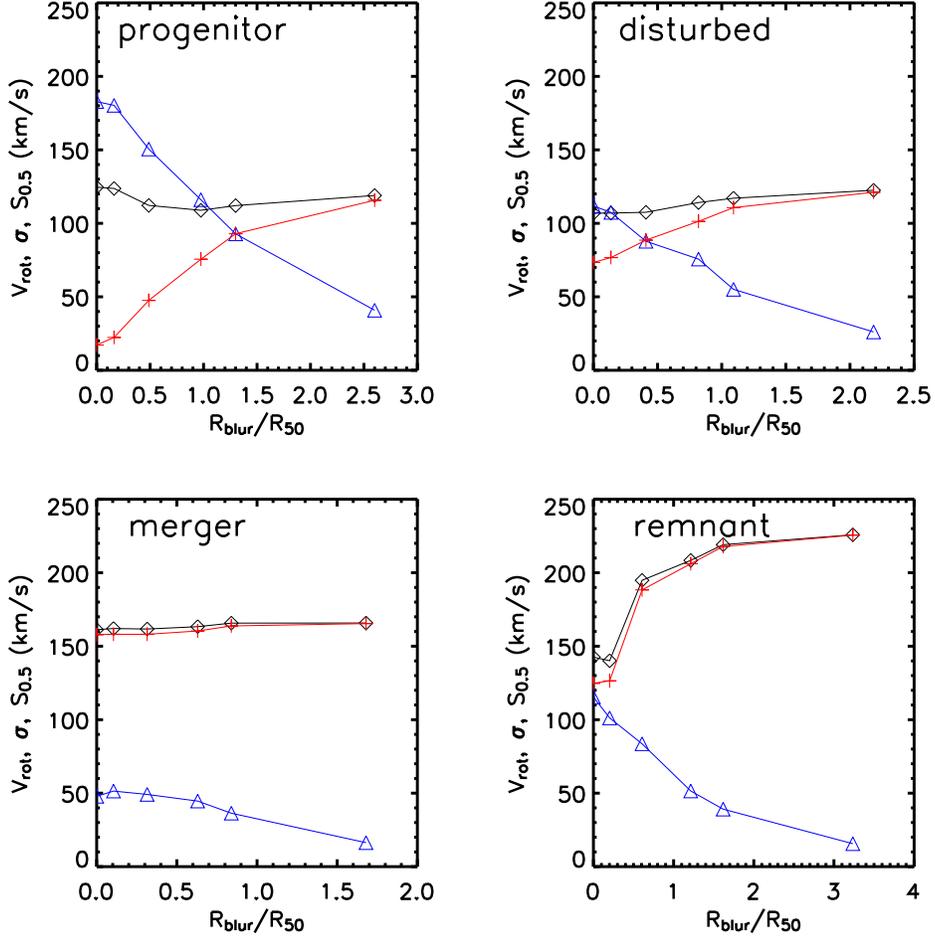}\\
\caption[Effect of blurring on kinematics.]{Effect of blurring on
  kinematics.  We take the four snapshots from our fiducial Sbc merger
  (undisturbed disk=top left, disturbed disk=top right, merger=bottom
  left, and remnant=bottom right) mock observed at $z=1$, and analyze the kinematics after
  blurring the particle positions by gaussians with $\sigma$'s of
  0.0'', 0.1'', 0.3'', 0.6'',0.8'', and 1.6''.  Blurring is normalized
  using the (unblurred) radius that contains 50\% of the stellar mass
  ($R_{\rm 50}$).  Lines and symbols are rotation velocity (blue
  triangles), $\sigma$ (red crosses), and ${\rm S_{0.5}}$ (black
  diamonds). Since the physical radii are different for each stage,
  and the blurring is for specific angular scales, each plot shows a different range
  of $R_{\rm 50}/R_{\rm blur}$.}
\label{fig:blur}
\end{center}
\end{figure}

\begin{figure}
\begin{center}
\includegraphics[width=0.64\textwidth]{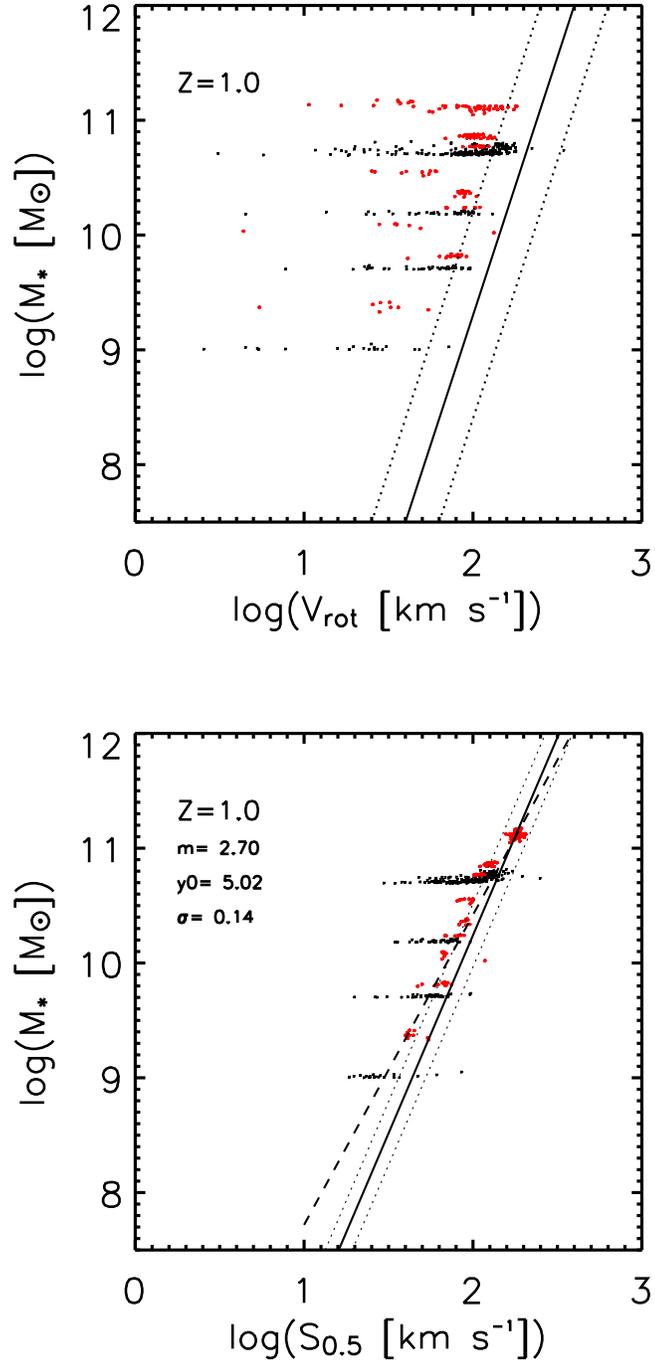}
\caption[The Stellar Mass TF and ${\rm S_{0.5}}$-Stellar Mass
relations with the analysis from the simulations done using only new
star particles.]{The Stellar Mass TF and ${\rm S_{0.5}}$-Stellar Mass
  relations analogous to Figure 2 with the analysis from
  the simulations done using only new star particles.}
\label{fig:ns}
\end{center}
\end{figure}

\begin{figure}
\begin{center}
\includegraphics[width=0.6\textwidth]{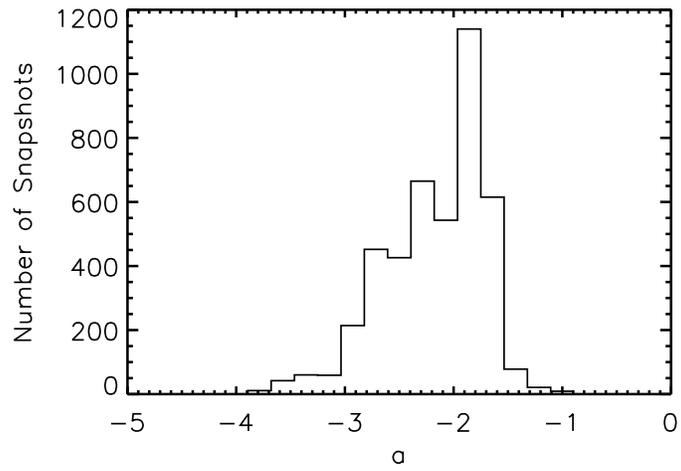}
\caption{Histogram of the power law slope fit to the density profiles near $R_{\rm 50}$ of each snapshot in the simulation set.}
\label{fig:a}
\end{center}
\end{figure}

\begin{figure}
\begin{center}
\includegraphics[width=0.6\textwidth]{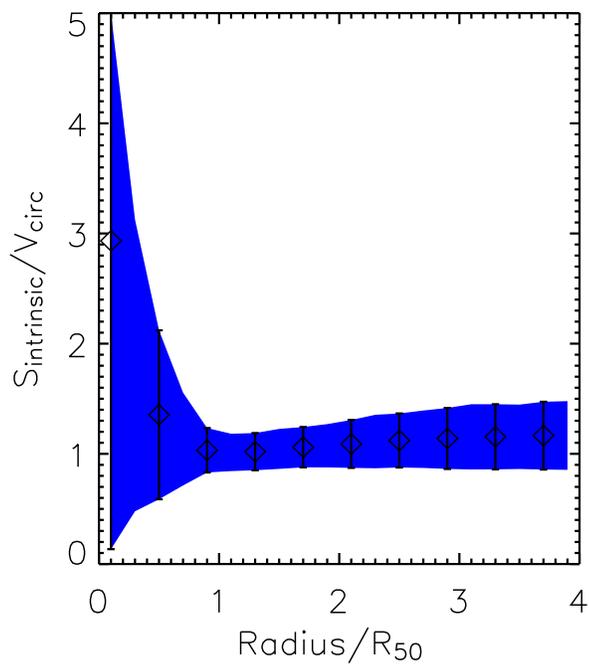}
\caption[Ratio of $S_{\rm intrinsic}$ to $V_{\rm circ}$ as a function of
radius]{Ratio of $S_{\rm intrinsic}$ and $V_{\rm circ}$ as a function of
  radius averaged over all simulation snapshots and normalized to
  $R_{50}$.  Blue range shows the $1\sigma$ scatter in the profiles.  At radii larger than $\sim R_{\rm 50}$, $S_{\rm intrinsic}$ is a good estimator of total enclosed mass.}
\label{fig:ratio}
\end{center}
\end{figure}

\begin{figure}
\begin{center}
\includegraphics[width=0.6\textwidth]{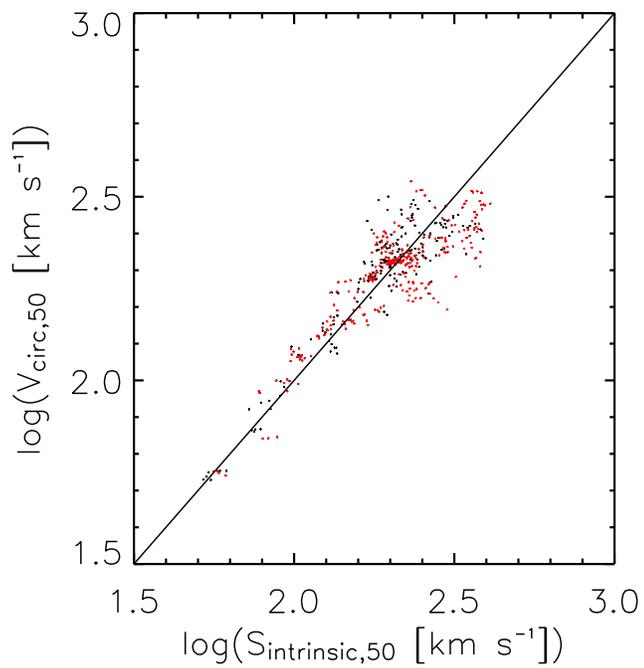}
\caption[Relation between intrinsic $S_{\rm intrinsic}$ and $V_{\rm
  circ}$ at $R_{\rm 50}$.]{Relation between intrinsic $S_{\rm
    intrinsic}$ and $V_{\rm circ}$ at $R_{\rm 50}$ as observed in the
  galaxy merger simulations.  Red points are cases for which both
  galaxies were analyzed together, most of which are remnants.  The
  $x=y$ line shows the rough equivalence of $S_{\rm intrinsic}$ and
  $V_{\rm circ}$.}
\label{fig:intr}
\end{center}
\end{figure}

\begin{figure}
\begin{center}
\includegraphics[width=0.6\textwidth]{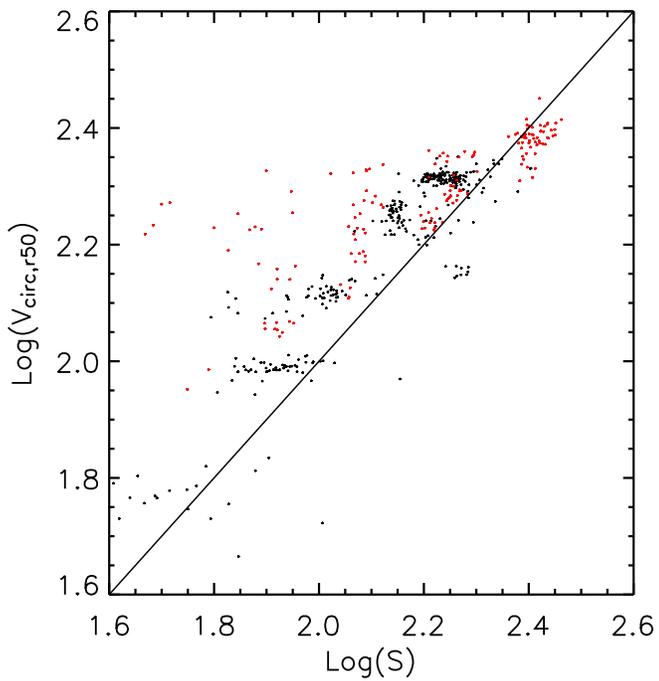}
\caption[Relation between $S_{\rm observed}$ and $V_{\rm circ}$ as
measured at $R_{\rm 50}$.]{Relation between $S_{\rm observed}$ and
  $V_{\rm circ}$ as measured at $R_{\rm 50}$.  Red points are cases for
  which both galaxies are analyzed together.  The line plotted is the
  x=y line, demonstrating the correspondence between the two values.}
\label{fig:vcirc}
\end{center}
\end{figure}

\end{document}